\documentclass[
amsmath,
prd,nofootinbib,floatfix,12pt,
]{revtex4}

\newcommand\beq{\begin{eqnarray}}
\newcommand\eeq{\end{eqnarray}}

\newcommand\stoponium{\eta_{\tilde t}}
\def\lsim{\mathrel{\rlap{\lower4pt\hbox{$\sim$}}
    \raise1pt\hbox{$<$}}}                
\def\gsim{\mathrel{\rlap{\lower4pt\hbox{$\sim$}}
    \raise1pt\hbox{$>$}}}            

\allowdisplaybreaks
\usepackage{graphicx}
\usepackage{setspace}
\usepackage{placeins}

\usepackage{dcolumn}
\usepackage{bm}

\begin{document}

\renewcommand{\theequation}{\arabic{section}.\arabic{equation}}
\renewcommand{\thefigure}{\arabic{section}.\arabic{figure}}
\renewcommand{\thetable}{\arabic{section}.\arabic{table}}

\title{\large \baselineskip=20pt 
LHC search for di-Higgs decays of stoponium and other scalars\\
in events with two photons and two bottom jets}

\author{Nilanjana Kumar$^1$ and Stephen P.~Martin$^{1,2}$}
\affiliation{
{\it $^1$Department of Physics, Northern Illinois University, DeKalb IL 60115} \\
{\it $^2$Fermi National Accelerator Laboratory, P.O. Box 500, Batavia IL 60510}
}


\begin{abstract}\normalsize \baselineskip=14pt 
We study the prospects for LHC discovery of a narrow resonance that decays 
to two Higgs bosons, using the final state of two photons and two bottom 
jets. Our work is motivated in part by a scenario in which two-body 
flavor-preserving decays of the top squark are kinematically forbidden. 
Stoponium, a hadronic bound state of the top squark and its anti-particle, 
will then form, and may have a large branching fraction into the two Higgs 
boson final state. We estimate the cross-section needed for a 5-sigma 
discovery at the 14 TeV LHC for such a narrow di-Higgs resonance, using 
the invariant mass distributions of the final state bottom jets and 
photons, as a function of the integrated luminosity. The results are also 
applicable to any other di-Higgs resonance produced by gluon fusion. 
\end{abstract}

\maketitle

\vspace{-0.3in}

\tableofcontents

\baselineskip=15.4pt

\section{Introduction\label{sec:intro}}
\setcounter{equation}{0}
\setcounter{figure}{0}
\setcounter{table}{0}
\setcounter{footnote}{1}

ATLAS \cite{ATLASHiggs,ATLAScombination} 
and CMS \cite{CMSHiggs,CMScombination} have confirmed the existence 
of a resonance with properties that are consistent with a 
minimal Standard Model Higgs scalar boson, $h$, 
with a mass near $126$ GeV. The precise value of $m_h$ is already  known
at roughly the 1\% level, and will surely improve in the future.
This provides an opportunity to 
search for new physics that lies beyond the Standard Model, 
by looking for new heavy particles that decay into $h$, exploiting 
the Higgs boson as a standard candle.

One such possibility is stoponium, $\stoponium$, a bound state of a top squark 
(stop) and its anti-particle. The stop will be stable enough to 
hadronize provided that it has no flavor-preserving two-body 
decays. The binding energy of the $J^{PC} = 0^{++}$ ground state 
of stoponium is of order a few GeV, and its width is typically about 
two orders of magnitude smaller.
It will decay primarily by annihilation into pairs of Standard Model 
particles, including final states $gg$, $WW$, $ZZ$, $hh$, $\gamma
\gamma$, $Z\gamma$, $t \overline t$, and $b \overline b$, as well as 
pairs of neutralinos, depending 
on the masses and the stop mixing angle and other 
supersymmetry-breaking parameters 
\cite{Drees:1993yr,Drees:1993uw,Martin:2008sv}. 
Therefore one can search for narrow invariant mass peaks of stoponium at 
the LHC or at future hadron colliders. 
The diphoton final state, as originally proposed in 
\cite{Drees:1993yr,Drees:1993uw} and studied more recently in
\cite{Martin:2008sv,Martin:2009dj,Younkin:2009zn} 
is a promising one due to its clean experimental 
signature and the excellent diphoton mass resolution of the 
LHC detectors. The $ZZ$ and $WW$ final states may also 
provide a viable discovery signature \cite{Barger:2011jt,Kim:2014yaa}.
Early work on stoponium at hadron colliders can be found in 
\cite{Moxhay:1985bg,Herrero:1987df,Barger:1988sp,Inazawa:1993qk}, and  discussions of
stoponium at linear colliders have been presented in
\cite{Gorbunov:2000tr,Gorbunov:2000nd,Fabiano:2001cw}.

If the stop mass is at least a few GeV larger than $m_h$, then the decay 
$\eta_{\tilde t} \rightarrow h h$ is kinematically allowed and 
also potentially observable \cite{Barger:1988sp}, and can easily have a 
branching ratio of tens of per cent. This possibility was explored in early work
for the case $2 m_h < m_{\stoponium} < m_W$ in ref.~\cite{Barger:1988sp}. 
In some more modern models, this decay can even have 
the dominant branching ratio if $m_{\stoponium}$ is not too far above 
the threshold $2 m_h$; see for example the model lines in Figure 8 in 
ref.~\cite{Martin:2008sv}, 
which illustrate cases with BR$(\stoponium \rightarrow hh) > 0.7$.
The BR$(\stoponium \rightarrow hh)$ tends to decrease slowly as
$m_{\stoponium}$ moves far above threshold.
The combination of the rare but clean decay 
$h\rightarrow \gamma\gamma$ and the high branching ratio decay $h 
\rightarrow b\overline b$ may provide the best opportunity to 
observe this mode. In this paper, we will therefore explore the ability 
of the LHC to discover stoponium through $pp \rightarrow \eta_{\tilde t} 
\rightarrow hh \rightarrow \gamma\gamma b \overline b$. This could 
either be an alternative discovery mode, or perhaps a confirmation of a 
discovery of stoponium in the $\stoponium \rightarrow \gamma\gamma$ or 
$\stoponium \rightarrow ZZ$ modes or of open stop pair production.

The stoponium state is produced through gluon fusion, as the 
near-threshold limit of open stop production. The production 
cross-section was computed through next-to-leading order (NLO) in 
ref.~\cite{Younkin:2009zn} in terms of the stoponium wavefunction at the 
origin. A resummed next-to-next-to-leading logarithm (NNLL) calculation 
is provided in \cite{Kim:2014yaa}; the effects of threshold resummation 
were found to be small. When needed, we will use the results of 
\cite{Younkin:2009zn} for convenience. The remaining uncertainties may 
well be dominated by the imperfect knowledge of the stoponium 
wavefunctions and production of the excited states. We note in 
particular that ref.~\cite{Younkin:2009zn} chose to include only the 1s 
and 2s stoponium states in the production cross-section. Although these 
give most of the production cross-section, there could be additional 
rate contributions
coming from production of higher excited states, if those decay to 
the $s$-wave states before decaying by annihilation.

More generally, the same signatures used to search for stoponium will 
apply to any narrow scalar di-Higgs resonance, including the heavier 
neutral Higgs scalar boson of the Minimal Supersymmetric Standard Model 
(MSSM), where there is sensitivity especially if $\tan\beta$ is not too 
large \cite{Plehn:1996wb,Baur:2003gp,Dolan:2012ac}, as well as other 
extensions of the Standard Model Higgs sector 
\cite{Liu:2013woa,No:2013wsa,Chen:2013emb}. The paper \cite{Chen:2013emb} 
contains a study similar to the present one, but with somewhat different 
motivations and procedures. A recent search by CMS \cite{CMSHhh} looks for 
$pp \rightarrow H \rightarrow hh$, 
and sets 95\% confidence level limits of order 5 pb on the production 
cross-section for $H$ masses below 360 GeV, but using channels other than 
$bb\gamma\gamma$. In another study by ATLAS \cite{ATLASbbbb} it is shown that 
a good sensitivity can be achieved for $m_{H} \geq 600$ while looking 
at resonances decaying via a pair of Higgs bosons to the $b \overline b b \overline b$ 
final state, with 19.5 fb$^{-1}$ of proton-proton collision data 
at $\sqrt{s} = 8$ TeV. In the rest of this paper, we will use $\eta$ to 
represent a generic di-Higgs resonance, although stoponium (denoted 
$\eta_{\tilde t}$) is our primary motivation. It should be noted that the 
signature for di-Higgs production is also used, with different kinematic 
requirements due to the non-resonant production, in order to study the 
trilinear Higgs self-coupling as a test of the Standard Model, for example 
see 
\cite{Glover:1987nx,Plehn:1996wb,Dawson:1998py,Baur:2002qd,Baur:2003gp,Dolan:2012rv,Papaefstathi,Baglio:2012np,Goertz:2013kp,Shao:2013bz,deFlorian:2013uza,Barr:2013tda,deFlorian:2013jea,Dolan:2013rja}. 
In the present paper this non-resonant Standard Model di-Higgs production 
is one of the backgrounds.

There are a variety of model-building motivations for light stops. For 
example, a light stop is required in the MSSM to enable weak-scale 
baryogenesis \cite{baryo}. A light stop scenario is also one way of 
accommodating the observed dark matter relic density \cite{WMAP,Planck} 
through efficient annihilations in the universe, if the lightest 
supersymmetric particle (LSP) is bino-like and 
$m_{\tilde{t_1}}-m_{\tilde{N_1}}$ is much smaller than the top quark 
mass, as the thermal abundance of DM can be reduced in such cases 
through stop-mediated neutralino annihilations and/or stop co-annihilations 
~\cite{stopcoannihilationA,stopcoannihilationB,Martin:2007gf,deSimone:2014pda}. 
The mass difference between the lighter stop and the LSP must be small 
enough to forbid flavor-preserving two-body decays in order to give the 
observed dark matter abundance. Finally, the naturalness arguments for 
``more minimal supersymmetry" \cite{Cohen:1996vb,Papucci:2011wy} 
generally incorporate light top squarks as a feature.

Recently, constraints on the light stop scenario have 
become available from ATLAS
\cite{TheATLAScollaboration:2013gha,TheATLAScollaboration:2013aia,Aad:2014qaa} and
CMS \cite{Chatrchyan:2013xna,CMS:2014yma}, ruling out 
significant parts of parameter space, including even
cases of stops that are nearly degenerate with the LSP.
However, there remain several holes in the exclusions,
including the cases 
$m_{\tilde t_1} - m_{\tilde N_1} \approx m_W + m_b$
and  
$m_{\tilde t_1} - m_{\tilde N_1} \approx m_t$.
Projected constraints by theorists 
reinterpreting other ATLAS and CMS searches claim 
\cite{Krizka:2012ah,Delgado:2012eu}
to fill in these holes up to about $m_{\tilde t_1} \approx 250$ GeV
(so $m_{\stoponium} \approx 500$ GeV), 
even using less than the full data sets of LHC Run 1. However, we prefer 
to take these exclusion
claims as preliminary until and unless they are confirmed by the experimental 
collaborations. Furthermore, if the stop decays as 
$\tilde t_1 \rightarrow jj$ through R-parity violation, where $j$
represents a light quark jet, then there are no exclusions at all
\cite{Evans:2012bf,Bai:2013xla} at present.
In this case, it may be that stoponium will be a competitive way 
to set model-independent limits on light stops for some time.
We will consider stoponium masses down to 275 GeV, corresponding to
top-squark masses down to about 138 GeV,
so that $\stoponium  \rightarrow hh$
is kinematically allowed.

\section{Event generation and simulation}
\setcounter{equation}{0}
\setcounter{figure}{0}
\setcounter{table}{0}
\setcounter{footnote}{1}

We used Madgraph 5 \cite{Alwall:2011uj} to generate events 
simulating $\eta$ production and decay, 
$pp \rightarrow \eta \rightarrow hh$, in proton-proton collisions 
at $\sqrt{s} = 14$ TeV. 
We used the model HEFT, an extension of the tree-level 
Standard Model to include an additional scalar, which we interpreted as $\eta$, 
and effective couplings $gg\eta$, $ggh$, and
$\gamma\gamma h$. We modified HEFT to also include a small
$\eta hh$ coupling to allow the decay of interest, which was then forced
at the level of event generation. 
The production cross-section for 
$pp \rightarrow \eta \rightarrow h h$ is 
taken as an input parameter, 
in order to maximize the generality of the results. We
set the Standard Model Higgs boson mass to be $m_h = 126$ GeV,
and used branching ratios BR$(h \rightarrow b\overline b) = 0.57$
and BR$(h \rightarrow \gamma\gamma) = 0.0022$.
 
In order to improve the statistics, we generated
signal events in which one of the 
$h$ was forced to decay to $b \overline b$ and the other to 
$\gamma\gamma$, and then normalized the resulting event 
sample according to the branching ratios and 
the assumed $pp \rightarrow \eta \rightarrow h h$ production rate just mentioned. 
We generated 100,000 events for each of $m_{\eta} = 
275$, 300, 325, 350, 375, 400, 425, 450, 475, 500, 525, 550, 575, 600, 650, 700,
800, 900, and 1000 GeV in this way.
All the signal samples as well as the background samples mentioned below were 
generated using Madgraph 5 and showered with Pythia 6 \cite{Sjostrand:2006za}. 

The possible backgrounds include non-resonant 
$\gamma\gamma b \overline b$ production, as well as
$\gamma\gamma c\overline{c}$ and 
$\gamma\gamma j(b/\overline{b})$ and
$\gamma\gamma j(c/\overline{c})$ and
$\gamma\gamma jj$ (where $j= g, u, d, s, \overline u, \overline d, \overline s$), and
$\gamma\gamma t\overline{t}$ and
$\gamma\gamma Z$ and
$t\overline{t}h$ and 
$Zh$ and 
$b\overline{b}h$ and
$hh$.
Production of the $hh$ background includes 
a triangular and a box diagram, 
but the effective coupling for the latter is not included in the 
version of HEFT we used. We therefore normalized the cross-section 
for the $hh$ background to be 40.2 fb, from \cite{Chen:2013emb}.
In the LHC detectors, electrons are sometimes misidentified as photons. We 
therefore included backgrounds from the processes $t\overline{t}$ (with 
two electrons faking photons) and $t\overline{t}\gamma$ (with one 
electron faking a photon). Here we used a probability of 0.0181 for each 
electron to fake a photon \cite{CMS12018}.
We did not include a possible 4-jet background ($jjjj$) 
because the efficiencies for two
jets to faking photons is very low, and the result
must also have two light-flavor jets mis-tagged as $b$-jets with
a rate of order $10^{-6}$, and this background tends to be distributed at 
low photon $p_T$ and invariant masses.
We did include backgrounds of the form $j\gamma b\overline {b}$, 
where one jet fakes a photon. Here, we used probabilities $1/20100$ for a 
gluon jet and $1/1680$ for a quark jet to fake a photon 
\cite{ATLAS2011007}. 

In order to obtain good statistics, we found it useful to 
put a generator-level cut on the 
minimum and maximum invariant mass of the diphoton pair 
($106<M_{\gamma\gamma}<146$) in the backgrounds listed above that
explicitly include $\gamma\gamma$, because a tighter cut will be imposed at the analysis
level anyway. For the $t\overline{t}h$ and 
$Zh$ and $b\overline{b}h$ backgrounds, we forced $h$ to decay 
to two photons, and for the $hh$ background we forced one $h$ to 
decay to
$\gamma\gamma$ and the other to decay to $b \overline b$, as for the 
signal. The event samples were normalized accordingly.

For the detector simulation 
we used Delphes 3 \cite{deFavereau:2013fsa}. We chose 
a conservative $b$-tagging efficiency for $b$-jets of 0.6. The 
efficiency of mistagging a charm as a
$b$-jet was taken to be 0.1, while for jets initiated by 
gluons and $u,d,s$ quarks the 
$b$-jet mistagging efficiency was chosen to be 0.001. 

\section{Event selection\label{sec:eventselection}}
\setcounter{equation}{0}
\setcounter{figure}{0}
\setcounter{table}{0}
\setcounter{footnote}{1}

In the analysis, we first selected events with exactly
two $b$-tagged jets and two photons. The leading and sub-leading 
(in transverse momentum, $p_T$) photon and $b$-jet 
are denoted $\gamma_1$, $\gamma_2$
and $b_1$, $b_2$, respectively. We then
applied cuts on the $p_T$, the pseudo-rapidity $\eta$ and $\Delta R \equiv 
\sqrt{(\Delta \eta) ^2 + (\Delta \phi)^2 }$) as follows, referred to below 
as event selection ${\bf S1}$:
\begin{itemize}
\item $p_T (b_1,b_2) > (40,30)$ GeV
\item $p_T (\gamma_{1}, \gamma_{2}) > (35,25)$ GeV 
\item $|\eta (b_{1},b_{2})| < 2.7$
\item $|\eta (\gamma_{1}, \gamma_{2})| < 2.5$
\item $\Delta R_{ij} > 0.5$, for $i,j = b_1,b_2,\gamma_1,\gamma_2$
\end{itemize}
The cuts on $b\overline{b}$ invariant mass, $p_T$ and $\Delta R$ 
has been chosen to retain most 
of the signal while reducing some major sources of background.
In particular, we found that reducing the $\Delta R$ cuts to 0.4 does not
increase the signal acceptance by a significant amount.
We performed the whole analysis with various 
other choices of leading and 
sub-leading $b$-jet $p_T$'s and found that other choices 
do not provide for a significantly better retention of signal over background.

\begin{figure}[!t]
\begin{minipage}[]{0.46\linewidth}
\includegraphics[width=7.5cm,angle=0]{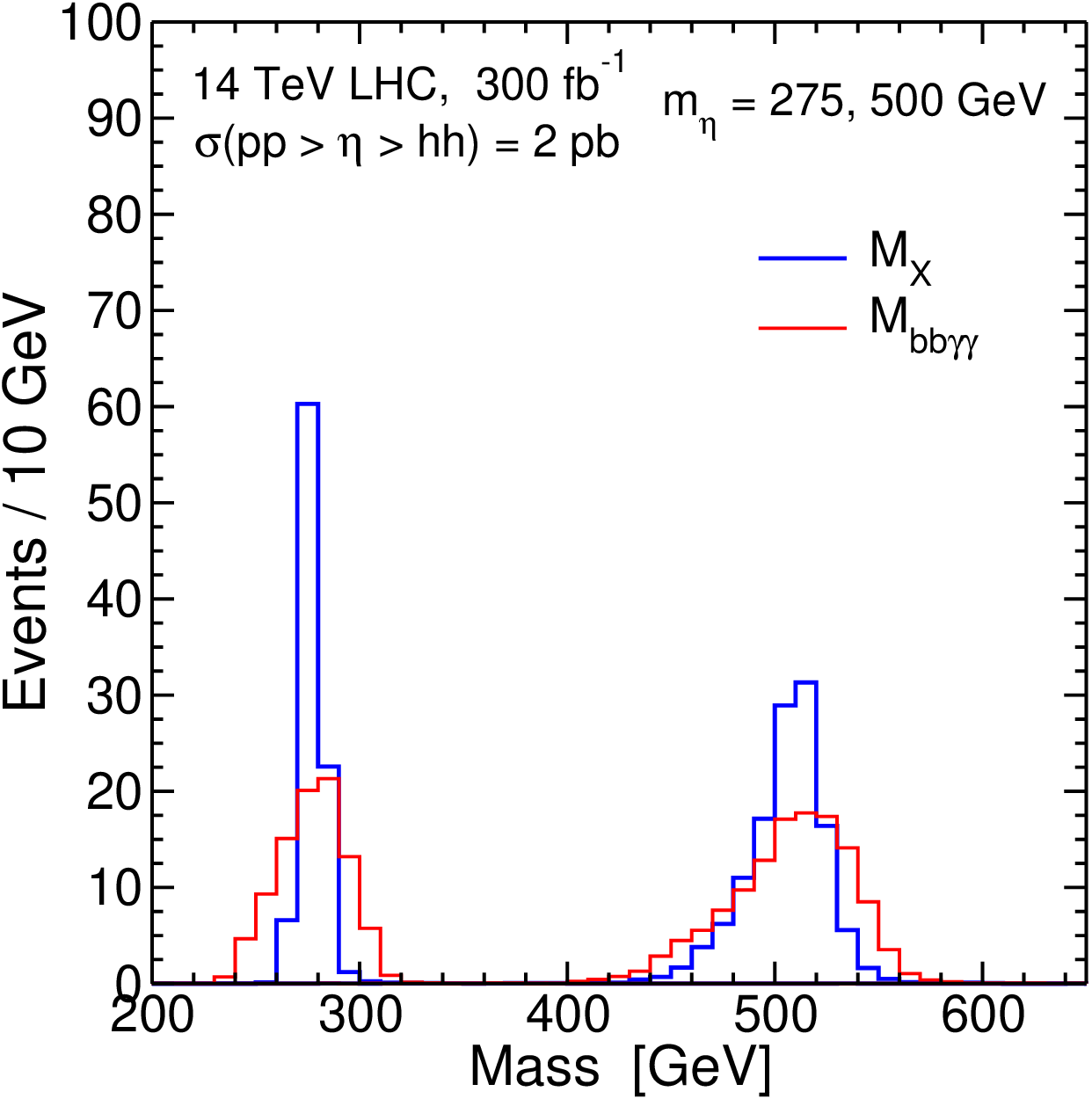}
\end{minipage}
\begin{minipage}[]{0.4\linewidth}
\phantom{xx}
\end{minipage}
\begin{minipage}[]{0.45\linewidth}
\caption{Distributions of $M_{bb\gamma\gamma}$ and $M_X$ as defined in 
eq.~(\ref{eq:defineMX}), for input masses $m_{\eta}=275$ GeV and $500$ GeV. 
Both distributions are based on
100,000 signal events $pp\rightarrow \eta \rightarrow hh$ with 
one $h$ forced to decay to $\gamma\gamma$ and the other to $b\overline 
b$, and with the distributions normalized by assuming 
$\sigma\times {\rm BR}(pp\rightarrow \eta \rightarrow hh)=2$ pb and 
an integrated luminosity of 300 fb$^{-1}$. 
The events were selected with the {\bf S3} 
cuts.  \label{fig:MX_M4}}
\end{minipage}
\end{figure}

\begin{figure}[!t]
\begin{minipage}[]{0.49\linewidth}
\includegraphics[width=8.2cm,angle=0]{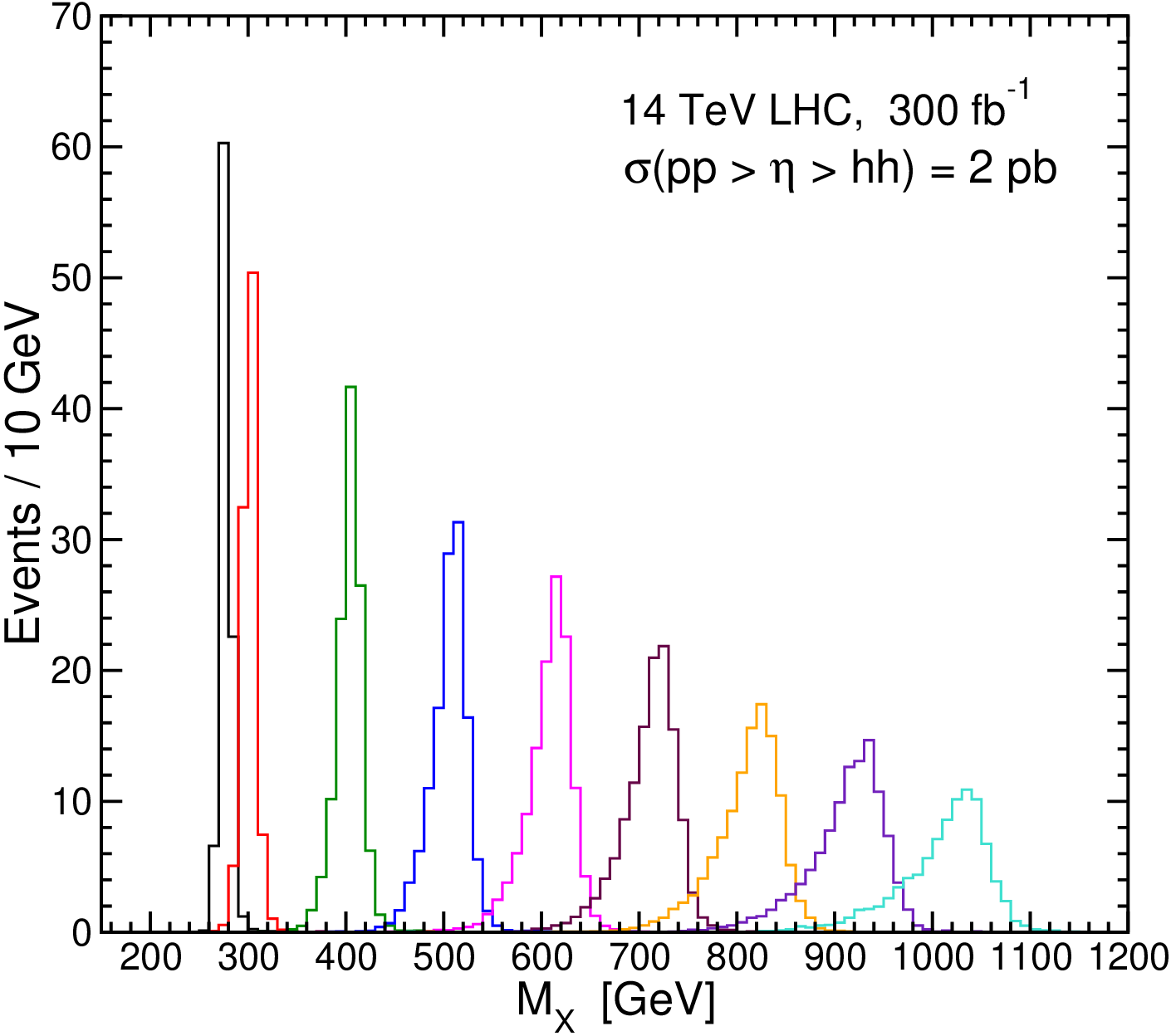}
\end{minipage}
\begin{minipage}[]{0.4\linewidth}
\phantom{x}
\end{minipage}
\begin{minipage}[]{0.45\linewidth}
\caption{\label{fig:MX_distributions_corr}Distributions of $M_X$ for $m_{\eta} = 275$, 
300, 400, 500, 600, 700, 800, 900, and 1000 GeV, 
for signal events, normalized assuming
$\sigma\times {\rm BR}(pp\rightarrow \eta \rightarrow hh)=2$ pb and 
an integrated luminosity of 
300 fb$^{-1}$, with the event selection {\bf S3} cuts imposed.} 
\end{minipage}
\end{figure}

Given the kinematics of the signal we are interested in, we then applied 
cuts on the invariant masses of the $\gamma\gamma$ pair, the $bb$ pair, 
and on the 4-body $\gamma\gamma bb$ system. For the last cut, we found 
that it is better to define a modified invariant mass $M_X$, according to
\beq
M_X \equiv M_{bb\gamma\gamma} - M_{bb} + m_h,
\label{eq:defineMX}
\eeq
where $m_h = 126$ GeV is the fixed, known Higgs mass. By subtracting
off $M_{bb}$ and adding in the true Higgs mass, one tends to mitigate 
the effects of $b$-jet momentum mismeasurements.
The distribution of $M_X$ has a sharper peak, and is 
concentrated closer to $m_\eta$, than the distribution of
$M_{bb\gamma\gamma}$. 
The sequence of event selection cuts we used is:
\begin{itemize}
 \item [{\bf S2}:] As in {\bf S1}, with $|M_{\gamma\gamma} - m_h| <$ 6 GeV,
 \item [{\bf S3}:] As in {\bf S2}, with $|M_{bb} - m_h| <$ 30 GeV, 
 \item [{\bf S4}:] As in {\bf S3}, with $|M_X-m_\eta| < 0.07 m_\eta$, where $m_\eta$ is 
            the position of the putative peak.
\end{itemize}
The widths of the $M_{\gamma\gamma}$ and $M_{bb}$ cuts are somewhat 
larger than the resolutions of a sample of single Higgs boson 
production, reflecting the performance we observed using Delphes when 
the Higgs bosons originate from heavy $\eta$ decays. Somewhat narrower 
(wider) windows could perhaps be used for smaller (larger) $m_\eta$, 
although we did not attempt to optimize this, since the optimization is 
likely to be quite different in real data than in our simulations. 
The advantage of using $M_X$ rather than the usual 4-body
invariant mass $M_{bb\gamma\gamma}$ is illustrated in Figure  
\ref{fig:MX_M4} for signal events that pass 
the {\bf S3} selection cuts, for $m_{\eta} = 275$ GeV
and for $m_{\eta} = 500$ GeV.
The distributions of $M_X$ as defined in eq.~(\ref{eq:defineMX}), 
for various different masses $m_{\eta}$ 
are shown in Figure \ref{fig:MX_distributions_corr}, again after the {\bf S3}
selection cuts. 
It can be seen that the $M_X$ distributions are peaked near the correct 
$\eta$ mass, and get wider as $m_{\eta}$ increases. For the larger 
values of $m_{\eta}$, especially above about 700 GeV,
the maximum of the $M_X$ distribution occurs 
somewhat above the true mass, but with a much fatter tail below than above. 
This is an effect that can be corrected for by the experimental 
collaborations in real data, and in our simulation most events are still 
within about $\pm 7$\% of the true value. Here, we expect that in 
practice a comparison between Monte Carlo simulations and an observed 
distribution will allow a hypothesis value of $m_\eta$ to be obtained in 
cases where a peak is present and large enough to possibly allow a 
5-sigma discovery claim. Given the luminosity requirements for a 
stoponium discovery, one may also expect that evidence for a stop, 
either in open production or in $\stoponium \rightarrow \gamma\gamma$ or 
$ZZ$ will have already accrued to allow for at least a rough estimate of 
the mass.

The fractions of $pp\rightarrow \eta \rightarrow hh$ signal events that pass 
selections {\bf S1}, {\bf S2}, {\bf S3}, and {\bf S4} are given in Table 
\ref{tab:signalfraction} for various values of $m_\eta$.
\begin{table}
\caption{The fraction of $pp\rightarrow \eta \rightarrow hh$ signal events 
at $\sqrt{s} = 14$ TeV
that pass selections {\bf S1}, {\bf S2}, {\bf S3}, and {\bf S4}. The
results were obtained for each $m_\eta$ by 
generating 100,000 events  $pp\rightarrow \eta \rightarrow hh$ 
with one $h$ forced to decay to
$\gamma\gamma$ and the other forced to decay to $b \overline b$, 
and then normalizing the results using
BR$(h \rightarrow \gamma\gamma) = 0.0022$ and 
BR$(h \rightarrow b \overline b) = 0.57$. \label{tab:signalfraction}}
\centering
\begin{center}
\begin{tabular}{|c|c|c|c|c|}
\hline
 $pp\rightarrow \eta \rightarrow hh$  & \multicolumn{4}{c|}{fraction $\times 10^{4}$}\\
\cline {2-5}
\hline
  $m_\eta$ (GeV) & {\bf S1} & {\bf S2} & {\bf S3} & {\bf S4}  \\
\hline
275 &	1.88 &1.80 &1.52 & 1.51\\
300 &	2.06 &1.97 &1.63 & 1.59\\
325 &	2.26 &2.13 &1.72 & 1.67\\
350 &	2.43 &2.23 &1.79 & 1.72\\
375 &	2.55 &2.30 &1.84 & 1.76\\
400 &	2.81 &2.48 &1.96 & 1.86\\
425 &	2.91 &2.49 &1.98 & 1.87\\
450 &	3.04 &2.52 &2.01 & 1.88\\
475 &	3.20 &2.60 &2.08 & 1.95\\
500 &	3.29 &2.63 &2.11 & 1.95\\
525 &  	3.36 &2.57 &2.08 & 1.92\\
550 &  	3.49 &2.60 &2.10 & 1.94\\
575 &	3.47 &2.53 &2.05 & 1.88\\
600 &	3.63 &2.59 &2.12 & 1.94\\
650 &   3.78 &2.53 &2.07 & 1.89\\
700 &   3.95 &2.52 &2.09 & 1.90\\
800 &   4.02 &2.32 &1.95 & 1.75\\
900 &   3.94 &2.14 &1.82 & 1.63\\
1000&   3.51 &1.84 &1.54 & 1.36\\
\hline
\end{tabular}
\end{center}
\end{table}
In order to obtain good statistics, the
results were obtained for each $m_\eta$ by 
generating 100,000 events $pp\rightarrow \eta \rightarrow hh$ with one $h$ forced to decay to
$\gamma\gamma$ and the other forced to decay to $b \overline b$, and then normalizing the results using
BR$(h \rightarrow \gamma\gamma) = 0.0022$ and BR$(h \rightarrow b \overline b) = 0.57$. The
nominal fraction of $pp\rightarrow \eta \rightarrow hh$ that will yield
$b\overline b \gamma\gamma$
before imposing any selection cuts and efficiencies 
is $2(0.0022)(0.57) = 0.00253$. After the {\bf S4} selection cuts,
the fraction of signal events surviving is of order $2 \times 10^{-4}$, 
and is largest for $m_\eta$ near 500 GeV.

\begin{figure}[!tb]
\begin{minipage}[]{0.45\linewidth}
\includegraphics[width=\linewidth,angle=0]{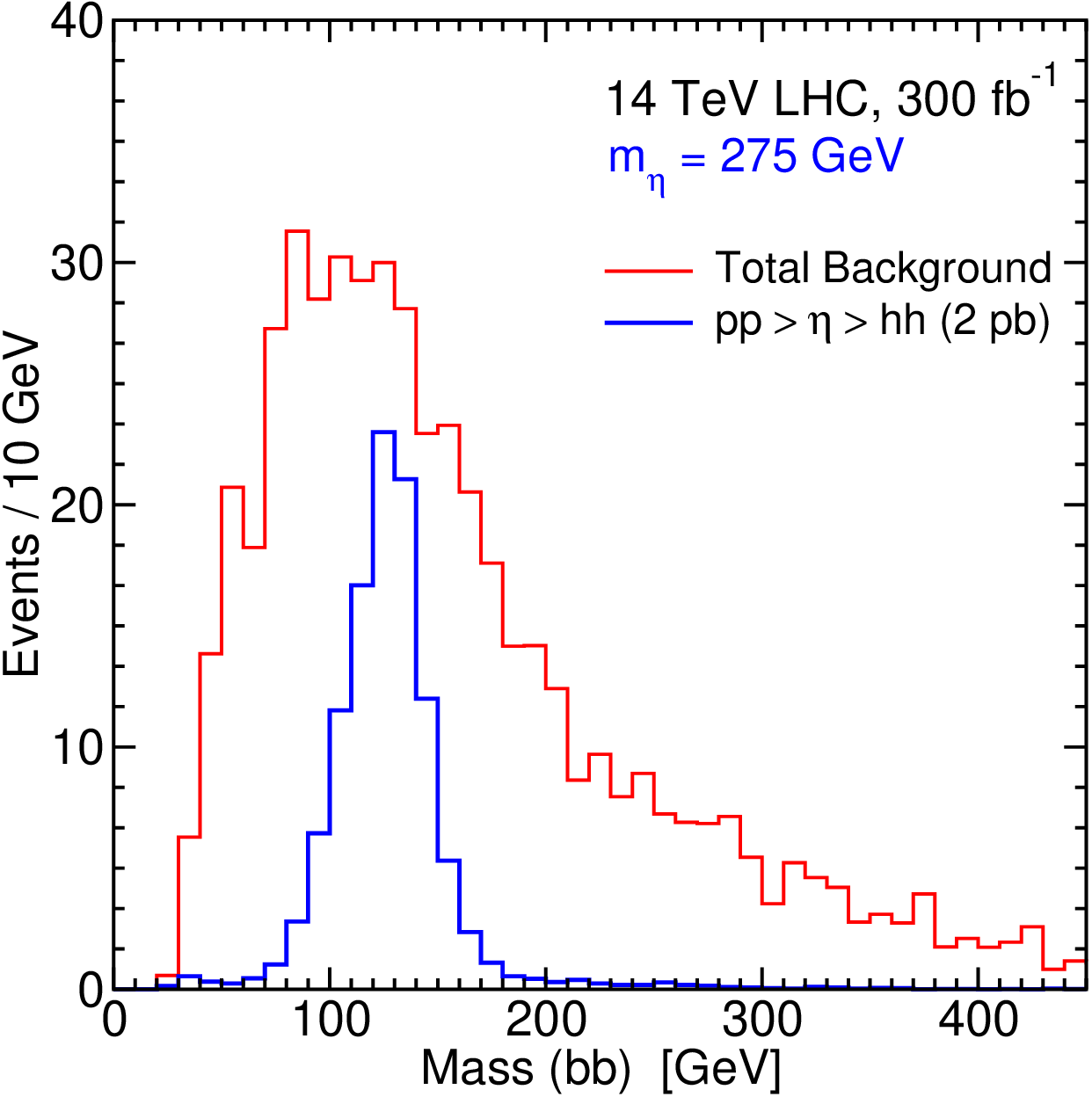}
\end{minipage}
\begin{minipage}[]{0.45\linewidth}
\begin{flushright}
\includegraphics[width=\linewidth,angle=0]{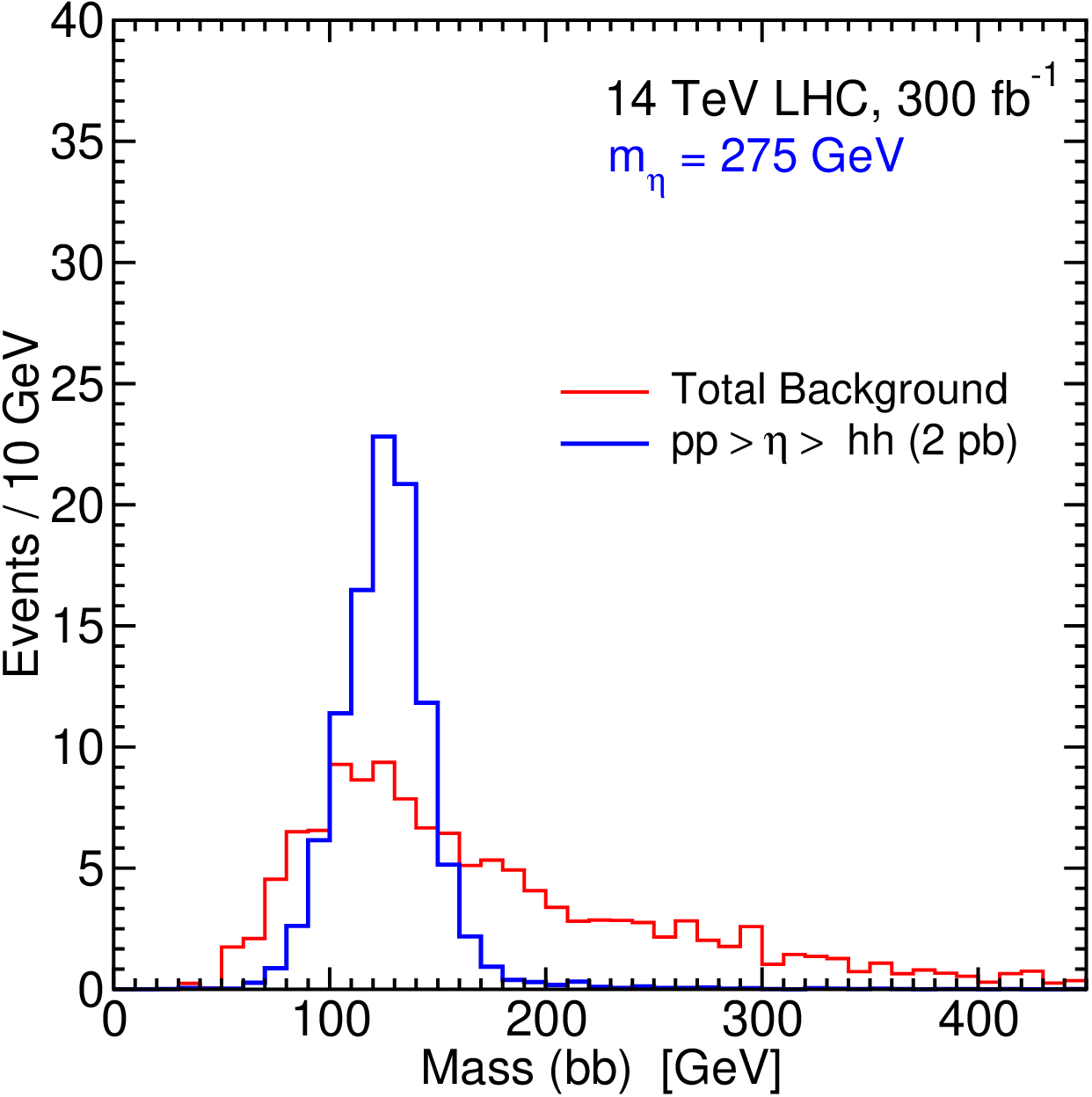}
\end{flushright}
\end{minipage}
\caption{\label{fig:Mbb} The signal and total background distributions of $M_{bb}$, 
after applying the {\bf S2} cuts (left panel) and 
after including in addition the 
{\bf S4} cut $|M_{X} - m_\eta| < 0.07 m_\eta$ (right panel), 
for $m_\eta = 275$ GeV. 
The normalizations assume 300 fb$^{-1}$ with 
$\sigma(pp \rightarrow \eta \rightarrow hh) = 2$ pb.}
\end{figure}

\begin{figure}[!tb]
\begin{minipage}[]{0.45\linewidth}
\includegraphics[width=\linewidth,angle=0]{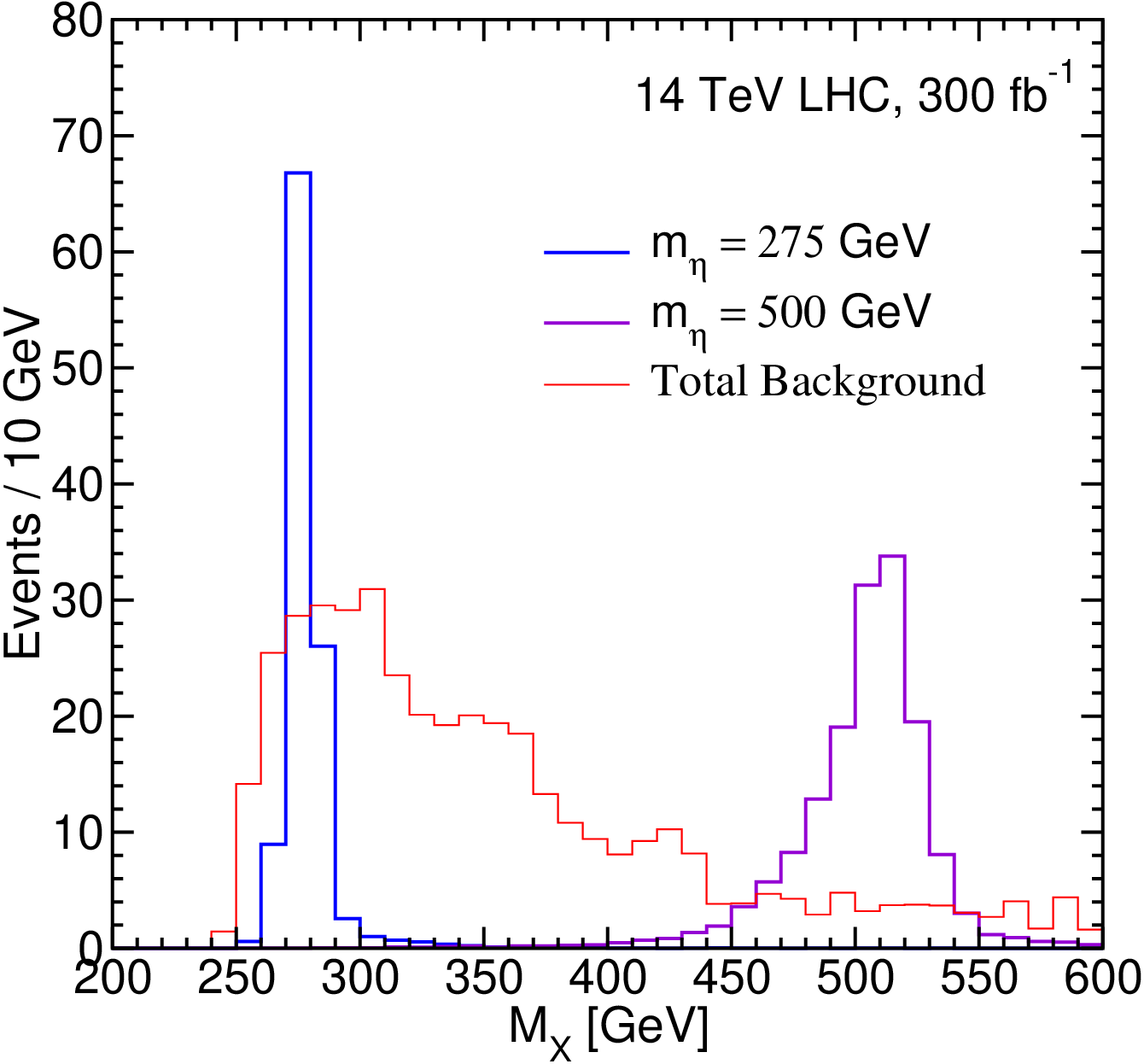}
\end{minipage}
\begin{minipage}[]{0.45\linewidth}
\begin{flushright}
\includegraphics[width=\linewidth,angle=0]{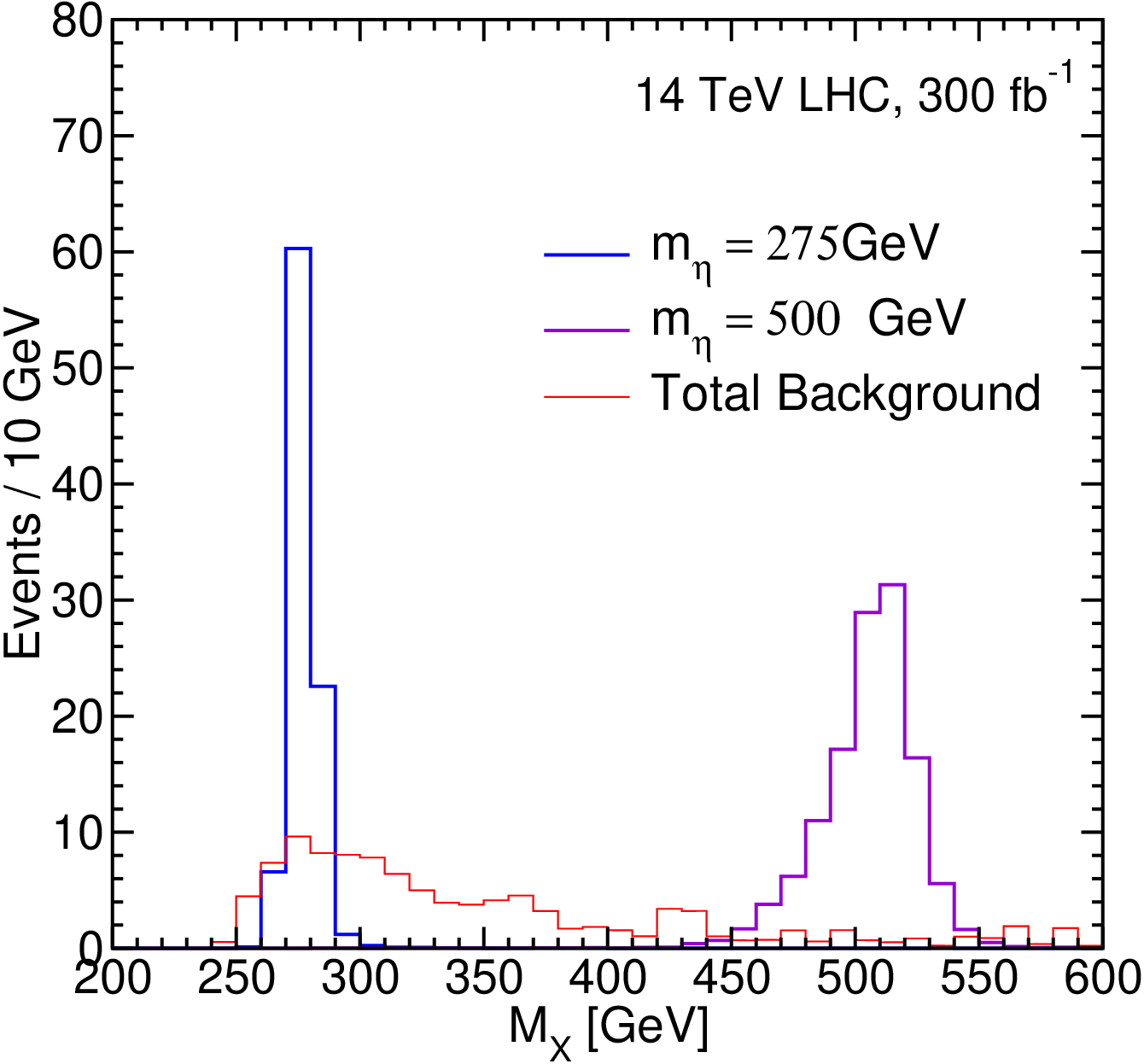}
\end{flushright}
\end{minipage}
\caption{The $M_X$ distributions of the total background and the signal 
are shown after event selections
{\bf S2} (left panel) and after {\bf S3} (right panel). For the signal, the distributions are shown for 
$m_\eta = 275$ GeV and 500 GeV, with
$\sigma(pp\rightarrow \eta \rightarrow hh) = 2$ pb in both cases.
The integrated luminosity is taken to be $300$ fb$^{-1}$.}
\label{fig:MX}
\end{figure}

\begin{figure}[!t]
\begin{minipage}[]{0.49\linewidth}
\includegraphics[width=7.5cm,angle=0]{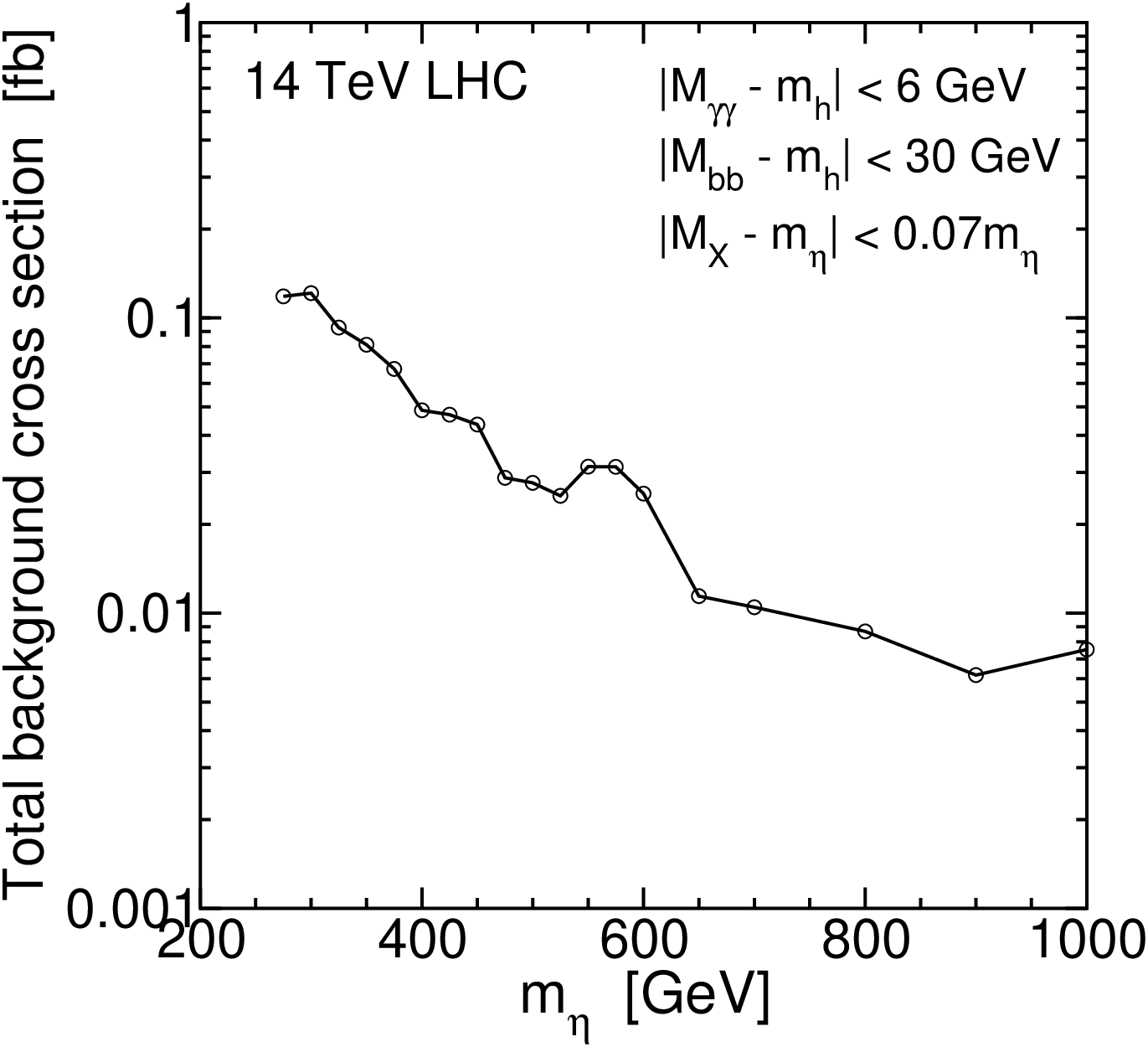}
\end{minipage}
\begin{minipage}[]{0.4\linewidth}
\phantom{x}
\end{minipage}
\begin{minipage}[]{0.45\linewidth}
\caption{\label{fig:totalback_corr} Total background cross section 
passing all cuts for event selection {\bf S4},
as a function of $m_\eta$, which enters into the $M_X$ cut.} 
\end{minipage}
\end{figure}

The backgrounds simulated and the cross-sections to pass the selections 
{\bf S1}, {\bf S2}, {\bf S3}, {\bf S4}, are shown in Table \ref{tab:back275}, for the
case that $m_{\eta} = 275$ GeV. (Only the {\bf S4} selection depends on the choice of $m_{\eta}$.)
\begin{table}
\caption{Significant background cross-sections after event 
selections {\bf S1}, {\bf S2}, {\bf S3} and {\bf S4}, 
for $m_{\eta}= 275$ GeV. 
The number of events generated, $N_{\rm gen}$, 
is also given. 
In order to improve statistics, 
the first seven backgrounds with $\gamma\gamma$ 
were generated with a cut $|M_{\gamma\gamma}-m_h|<20$ GeV, 
while the next four backgrounds
were generated with $h \rightarrow \gamma\gamma$ forced, 
and the $hh$ background was generated
with one $h$ forced to decay to $\gamma\gamma$ 
and the other to $b \overline b$.}
\label{tab:back275}
\begin{center}
\vspace{-0.1in}
\begin{tabular}{|c|c|c|c|c|c|}
\hline
~~~Background~~~ &  $N_{\mbox{gen}}$ & \multicolumn{4}{c|}{$\sigma_{\mbox{pass}}$ (fb)}\\
 \cline {3-6}
  && {\bf S1} & {\bf S2} & {\bf S3} & {\bf S4}  \\
\hline
$pp\rightarrow \gamma\gamma b\overline{b}$ & 200000 & 0.944 & 0.284 & 0.0861 & 0.0329
\\
$pp\rightarrow \gamma\gamma c\overline{c}$ & 440000 & 0.303 & 0.0912 & 0.0301 & 0.0131
\\
$pp\rightarrow \gamma\gamma t\overline{t}$ & 200000 & 0.119 & 0.0640 & 0.0176 & 0.00449
\\
$pp\rightarrow \gamma\gamma j(b/\overline{b})$ & 200000 & 0.764 & 0.233 & 0.0818 & 0.0217
\\
$pp\rightarrow \gamma\gamma j(c/\overline{c})$ & 600000 & 0.369 & 0.114 & 0.0337 & 0.0078
\\
$pp\rightarrow \gamma\gamma jj$ & ~1200000~ & 0.540 & 0.186 & 0.0723 & 0.0723
\\
$pp\rightarrow \gamma\gamma Z$  & 200000 & ~0.0462~ & ~0.0172~ & ~0.00220~ & ~0.00052~
\\
$pp\rightarrow t\overline{t} h$ & 100000 & 0.0733 & 0.0631 & 0.0171 & 0.00413
\\
$pp\rightarrow Zh$  & 100000 & 0.00919 & 0.00792 & 0.00329 & 0.00066 
\\
$pp\rightarrow b\overline{b}h$ & 100000 & ~0.0113 & ~0.00992~ & ~0.00251~ & ~0.00052~ 
\\
$pp\rightarrow hh$  & 100000 & 0.00927 & 0.00838 & 0.00682 & 0.00212
\\
$pp\rightarrow tt$  & 500000 & 0.108 & 0.00748 & 0.00216 & 0.00090
\\
$pp\rightarrow \gamma tt$ & 500000 & 0.157 & 0.00992 & 0.00267 & 0.00086
\\
$pp\rightarrow g \gamma b \overline{b}$ & 500000 & 0.3522 & 0.0314 & 0.0113 & 0.00411
\\
$pp\rightarrow (q/\overline q)\gamma b \overline{b}$ & 500000 & 3.568 & 0.253 & 0.0763 & 0.0173
\\
\hline
Total & & 7.374 & 1.379 & 0.446 & 0.118 \\
\hline
\end{tabular}
\end{center}
\end{table}
In Figure \ref{fig:Mbb}, we show for $m_\eta = 275$ GeV the $M_{bb}$ distributions 
for the signal and the background after
applying the selections {\bf S2}, and again after 
including the {\bf S4} cut on $M_X$. 
The latter cut is seen to strongly reduce the 
background while keeping most of the signal.
In Figure \ref{fig:MX} we show
the $M_X$ distributions for the total background and for the signal, 
assuming $\sigma(pp\rightarrow \eta \rightarrow hh) = 2$ 
pb, for two choices $m_\eta = 275$ and 500 GeV. The left panel
shows the $M_X$ distributions after the event selections {\bf S2}, and the right panel
after including the {\bf S3} selection cut on $M_{bb}$, which clearly 
helps to give a good
discrimination against total background. 
These distributions are again shown weighted according to
300 fb$^{-1}$ integrated luminosity. 
Because the event selection {\bf S4} cut depends on the $m_\eta$ of the 
putative peak, the background drops significantly with higher masses. 
This is shown in Table \ref{tab:backgroundsM} for $m_\eta = 300$, 400, 
500, 600, 700, 800, 900, 1000 GeV. Note that for smaller $m_{\eta}$, the 
backgrounds are largest for $\gamma\gamma b\overline b$ and 
$\gamma\gamma j (b/\overline{b})$ and $j\gamma b \overline b$, but for higher $m_{\eta}$ we find that 
the largest background is $\gamma\gamma jj$ for $j = g,u,d,s,\overline u, \overline d,\overline s$. 
Clearly these results will be dependent on the ability of the detector 
analyses to minimize mis-tags of gluon and light quark jets as $b$-jets and photons.
\begin{table}
\caption{Background cross sections in fb after selections {\bf S4}, 
for $m_\eta = 300$, 400, 500, 600, 700, 800, 900, 1000 GeV. Cases where no events passed the {\bf S4} selections
are listed with $\leq$ and the 1-event cross-section of our sample. 
For these and other values of $m_\eta$, the total backgrounds after the {\bf S4} cuts are shown in Figure \ref{fig:totalback_corr} below.
\label{tab:backgroundsM}}
\begin{center}
\vspace{-0.2in}
\begin{tabular}{|c|c|c|c|c|c|c|c|c|}
\hline
 Background & \multicolumn{8}{c|}{$\sigma_{\mbox{pass}}$ (fb) for various $m_{\eta}$ in GeV }\\
 \cline {2-9}
  & 300 & 400  & 500 & 600  & 700 & 800 & 900 & 1000 \\
\hline
$\gamma\gamma b\overline{b}$  & 0.0291 & 0.00797 & 0.00286 & 0.00082 & 0.00061 & 0.00020 &  0.00041 &  0.00010\\
$\gamma\gamma c\overline{c}$  & 0.00921 & 0.00146 & 0.00048 & 0.00048 & $\leq 0.00048$ & $\leq 0.00048$ & $\leq 0.00048$ & $\leq 0.00048$\\
$\gamma\gamma t\overline{t}$  & 0.00497 & 0.00253 & 0.00104 & 0.00045 & 0.00016 & 0.00010 & 0.00003 & 0.00001\\  
$\gamma\gamma j(b/\overline{b})$  & 0.0199 & 0.00938 & 0.00563 & 0.00338 & 0.00525 & 0.00263 & 0.00075 & 0.00037\\
$\gamma\gamma j(c/\overline{c})$ & 0.01037 & 0.00648 & 0.00389 & 0.00130 & $\leq 0.00130 $ &  $\leq 0.00130$ & $\leq 0.00130 $ & $\leq 0.00130 $\\
$\gamma\gamma jj$ & 0.01446 & 0.00482 & 0.00482 & 0.0121 & $\leq 0.00241$ & 0.00241 & 0.00241 & 0.00482\\ 
$\gamma\gamma Z$ & ~0.00036~ & ~0.00040~ & ~0.00016~ & 0.00012 & 0.00008 & 0.00012 & $\leq 0.00004$ & $\leq 0.00004$ \\
$t\overline{t}h$ & 0.00483 & 0.00255 & 0.00088 & 0.00045  & 0.00024 & 0.00006 & 0.00005 & 0.00002\\
$Zh$ & 0.00066 & 0.00055 & 0.00033 & 0.00018 & 0.00011 & 0.00006 & 0.00002 & 0.00001\\
$b\overline{b}h$  & 0.00050 & 0.00037 & 0.00023 & 0.00013 & 0.00010 & 0.00007 & 0.00003 & 0.00004\\
$hh$ & 0.00208 & 0.00080 & 0.00032 & 0.00015 & 0.00005 & 0.00003 & 0.00002 & 0.000004\\
$tt$ & 0.00091 & 0.00011 & 0.00002 & $\leq 3\!\! \times\!\! 10^{-6} $ & $\leq 3\! \times\! 10^{-6} $ & $\leq 3\! \times\! 10^{-6} $ & $\leq 3\! \times\! 10^{-6} $ & $\leq 3\! \times\! 10^{-6} $ \\
$\gamma tt$ & 0.00090 & 0.00028 & 0.00005 & 0.000025 & 0.000035 & 0.000015 & $\leq 5\! \times\! 10^{-6} $ &$\leq 5\! \times\! 10^{-6}$\\
$g\gamma b \overline{b}$ &0.00412  &0.00091 & 0.00030 & 0.00011 &0.00004 &$\leq 0.00004 $ &$\leq 0.00004 $ &$\leq 0.00004 $ \\
$(q/\overline{q}) \gamma b \overline{b}$ & 0.0187 & 0.0101 & 0.00662 & 0.00576 & 0.00201 & 0.00115 & 0.00058 & 0.00029\\
\hline
Total & 0.1213 & 0.0487 & 0.0276 & 0.0254 & 0.0105 & 0.0087 & 0.0062 & 0.0075 \\
\hline
\end{tabular}
\end{center}
\end{table}
\FloatBarrier
The results for the total background cross-sections passing events selection {\bf S4}, 
as a function of $m_\eta$,
are plotted in Figure \ref{fig:totalback_corr}.

\section{Discovery prospect projections for the 14 TeV LHC}
\setcounter{equation}{0}
\setcounter{figure}{0}
\setcounter{table}{0}
\setcounter{footnote}{1}

In actual experimental data, the appearance of a peak in the $M_X$ 
distribution would allow a discovery if it is large enough. The 
background levels should be determined with some accuracy from data, due 
to the presence of several sideband control regions. These include 
events with $M_{\gamma\gamma}$ outside of the window specified in the 
{\bf S2} cut, events with $M_{bb}$ outside of the window specified in 
the {\bf S3} cut, and events with $M_X$ outside of the window specified 
in the {\bf S4} cut. We therefore assume that the determination of 
backgrounds for the search will be mostly statistical, and set a 
requirement for a 5-sigma observation of the signal by demanding that 
$S/\sqrt{B} > 5$, where $S$ and $B$ are the numbers of signal and 
background events, respectively, that pass the {\bf S4} selection. While 
this does not account for the ``look-elsewhere" effect, it is likely 
that because of the large luminosities required, by the time a stoponium 
discovery search becomes relevant, there will be other evidence either 
from one or both of the channels $\stoponium \rightarrow \gamma\gamma$ 
or $\stoponium \rightarrow ZZ$ or from open stop production, or perhaps 
from stops obtained from gluino decays. We also require a minimum of $S>10$ 
signal events for a discovery, which becomes important when the signal 
and background cross-sections are both low.

\begin{figure}[!t]
\begin{minipage}[]{0.49\linewidth}
\includegraphics[width=8.2cm,angle=0]{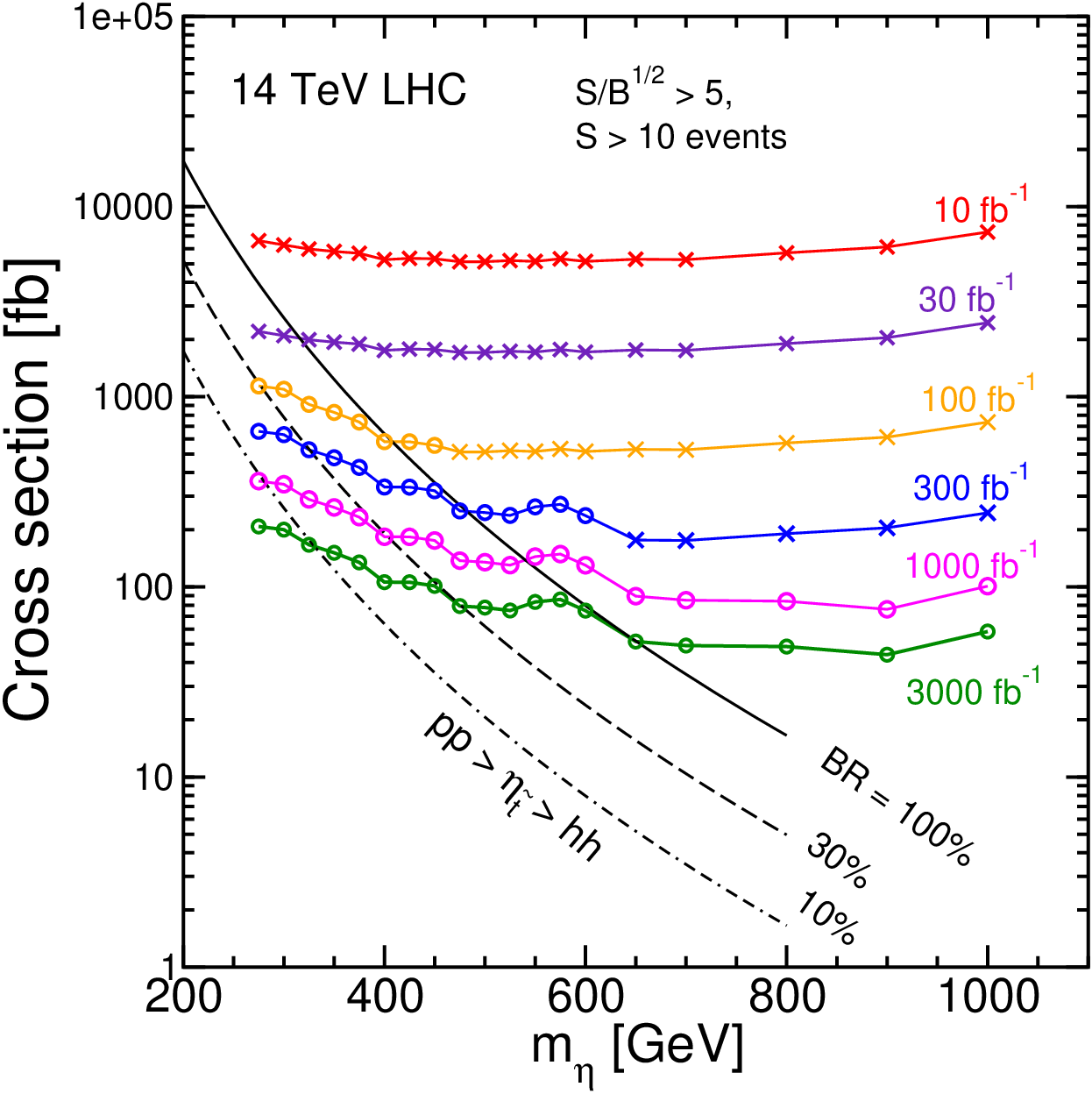}
\end{minipage}
\begin{minipage}[]{0.4\linewidth}
\phantom{x}
\end{minipage}
\begin{minipage}[]{0.45\linewidth}
\caption{\label{fig:sigmadiscovery} The $\sigma(pp\rightarrow \eta 
\rightarrow hh)$ cross sections needed for an expected $S/\sqrt {B}>5$, 
$S>10$ event discovery as a function of $m_\eta$, for integrated 
luminosities 10, 30, 100, 300, 1000, and 3000 fb$^{-1}$ in $pp$ 
collisions at $\sqrt{s} = 14$ TeV. The points marked with a circle are 
those that have an expected $S/\sqrt{B} = 5$ with $S > 10$ events, while 
those marked with an X have an expected $S = 10$ events with $S/\sqrt{B} 
> 5$. Also shown are the predicted cross-sections for stoponium 
production, $\sigma (pp\rightarrow \stoponium \rightarrow hh)$, based on 
ref.~\cite{Younkin:2009zn} for NLO $\sigma (pp\rightarrow \stoponium)$ 
and with assumed BR$(\stoponium \rightarrow hh) = $ 100\%, 30\%, 10\%.}
\end{minipage}
\end{figure}

\begin{figure}[!t]
\begin{minipage}[]{0.49\linewidth}
\includegraphics[width=8.2cm,angle=0]{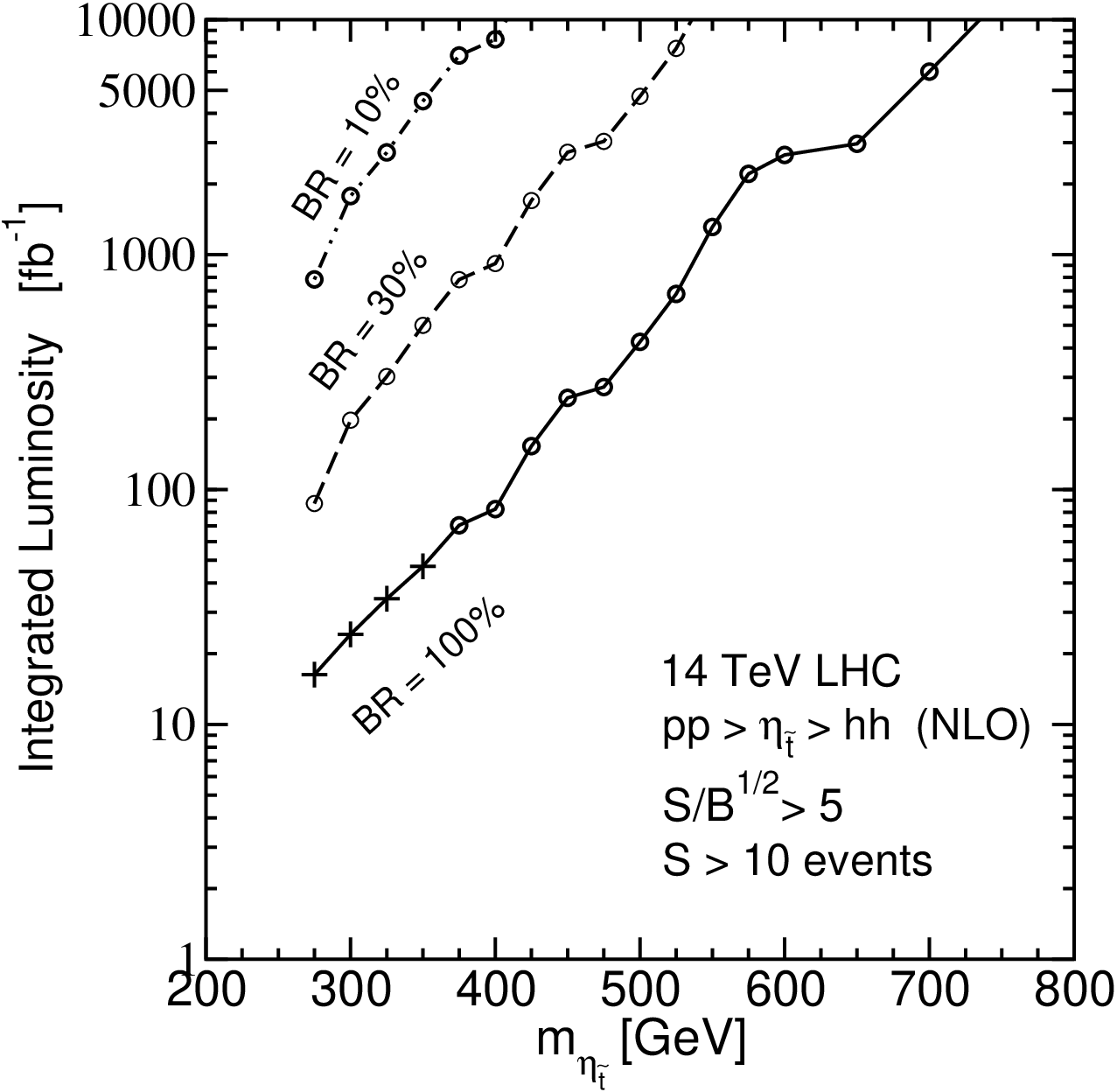}
\end{minipage}
\begin{minipage}[]{0.4\linewidth}
\phantom{x}
\end{minipage}
\begin{minipage}[]{0.45\linewidth}
\caption{\label{fig:L_5sigma}Total integrated luminosity needed for an 
expected $S/\sqrt {B} >5$ and $S > 10$ events, as a function of 
$m_{\stoponium}$, for $pp\rightarrow \stoponium \rightarrow hh$ at 
$\sqrt{s} = 14$ TeV, taking the NLO cross-section for $pp\rightarrow 
\stoponium$ from ref.~\cite{Younkin:2009zn} and assuming 100\%, 30\%, 
and 10\% branching ratios for $\stoponium \rightarrow hh$. The points 
marked with a circle have an expected $S/\sqrt {B} =5$ and $S > 10$ 
events, while those marked by a $+$ symbol have an expected $S = 10$ 
events and $S/\sqrt {B} >5$.}
\end{minipage}
\end{figure}

In Figure \ref{fig:sigmadiscovery} we show the cross-section $\sigma 
(pp\rightarrow \eta \rightarrow hh)$ needed for $S/\sqrt{B} > 5$ and $S 
> 10$, as a function of $m_\eta$, for various integrated luminosities 
and $\sqrt{s} = 14$ TeV. 
We see that with an integrated luminosity of 
100 fb$^{-1}$ at the 14 TeV LHC, one should be able to discover (or, with the 
look-elsewhere effect, provide strong evidence for) the resonant process 
$pp \rightarrow \eta \rightarrow h h$, provided the cross-section 
exceeds 500 fb to 1.2 pb, depending on the mass. 
Put another way, a di-Higgs resonance with a cross-section for $pp 
\rightarrow \eta \rightarrow hh$ of 1.2 pb can be easily 
discovered with less than 100 fb$^{-1}$ of integrated luminosity, independent 
of its mass as long as it is larger than about 275 GeV. 
With 300 fb$^{-1}$, 
it may be possible to discover a di-Higgs resonance with a cross-section 
as low as 175-250 fb, if its mass is in the 600-1000 GeV range, although 
this is limited by statistics. 
However, for the specific 
case of stoponium, the 
expected cross-sections fall very steeply with mass.  
For comparison, also shown in Figure \ref{fig:sigmadiscovery} 
are the predicted cross-sections for stoponium production, $\sigma 
(pp\rightarrow \stoponium \rightarrow hh)$, based on 
ref.~\cite{Younkin:2009zn} for $\sigma (pp\rightarrow \stoponium)$ and with 
assumed BR$(\stoponium \rightarrow hh) = $ 100\%, 30\%, and 10\%, as 
indicated.
Figure \ref{fig:L_5sigma} shows the integrated luminosity 
required for discovery of 
stoponium as a function of $m_{\stoponium}$, for $100\%$, $30\%$, and 
$10\%$ branching ratios of $\stoponium$.
With as little as 17 fb$^{-1}$ at $\sqrt{s} = 14$ TeV, the LHC could 
be able to discover the di-Higgs decay of stoponium with 
$m_{\stoponium} = 275$ GeV, if the branching ratio for $\eta_{\stoponium} \rightarrow hh$ is close to 100\%. 
However, even in this optimistic branching ratio case, the discovery 
potential with 300 fb$^{-1}$ 
runs out for stoponium masses heavier than about 500 GeV, 
corresponding to a 250 GeV top squark. For lower branching ratios, the 
required integrated luminosity is clearly much higher.

\section{Outlook\label{sec:outlook}}
\setcounter{equation}{0}
\setcounter{figure}{0}
\setcounter{table}{0}
\setcounter{footnote}{1}

In this paper we have examined the prospects of detecting stoponium and 
other di-Higgs resonances in the $b\overline{b}\gamma\gamma$ channel at 
the LHC with $\sqrt{s} = 14$ TeV. Our results outlined in the previous section
can be compared with the 
heavy Higgs search projections using the same final state made in 
ref.~\cite{Chen:2013emb}, which we became aware of while the present 
work was in progress. Ref.~\cite{Chen:2013emb} used a somewhat different 
set of analysis parameters, including a higher $b$-tagging efficiency of 
0.7 compared to our more conservative 0.6, a significantly smaller 
$M_{bb}$ window, and various other different choices for cuts. 
Nevertheless, comparing our results to Table III of 
ref.~\cite{Chen:2013emb} for the case of a 300 GeV scalar, we find a 
quite similar projection for the $S/\sqrt{B}$. Other results in 
ref.~\cite{Chen:2013emb} are based on the particular $(\alpha,\beta)$ 
parameter space of two Higgs doublet models, so that direct comparisons 
are difficult for other mass cases. Our work is therefore complementary to 
ref.~\cite{Chen:2013emb} in the sense that we presented our projections 
without tying to a specific model for the production cross-section.

In this paper, we did not attempt to make projections for the ability of 
the LHC to produce 95\% confidence level exclusions for stoponium or 
other di-Higgs resonances, which will be appropriate in the case of an 
absence of any significant candidate peaks in the $bb\gamma\gamma$ 
invariant mass distribution. To do this will require more sophisticated 
analyses techniques, rather than just simple cuts. However, clearly the 
sensitivity of the LHC to making exclusions should be considerably 
stronger than the discovery projections made here. Besides the 
$bb\gamma\gamma$ final state looked at here, other channels with higher rates
are worthy of consideration \cite{Plehn:1996wb}-\cite{CMSHhh}. 
In any case, it should be clear on general grounds that LHC searches for 
di-Higgs resonances should be a priority in the future, in order to 
exploit the Higgs discovery as a possible window to new physics.

Note added: after this paper appeared, the ATLAS collaboration released
the results \cite{Aad:2014yja} for searches for resonant and 
non-resonant $hh$ production in the
$\gamma\gamma bb$ final state, with $\sqrt{s} = 8$ TeV. The 95\% exclusion on 
the cross-section at $\sqrt{s} = 8$ TeV varies from 800 to 3500 fb when 
the resonance mass is less than 500 GeV, 
and is weaker than expected for some resonance masses below 350 GeV.

\noindent {\it Acknowledgments:} We thank Jahred Adelman and an anonymous 
referee for useful comments, and 
Chul Kim, Ahmad Idilbi, Thomas Mehen, 
and Yeo Woong Yoon for communications regarding
the stoponium production cross-section calculation of ref.~\cite{Kim:2014yaa}.
This work was supported in part by the National
Science Foundation grant number PHY-1068369. 


\end{document}